\documentclass[11pt,a4paper]{article} 
\usepackage{epsfig} 

\title{The Baby Skyrme Models and Their Multi-Skyrmions}  \author{Tom Weidig,
\\Centre for Particle Theory,\\Mathematical Sciences,  \\ University of
Durham\\email: tom.weidig@durham.ac.uk} 

\newcommand{\ph}{\vec{\phi}} \newcommand{\pha}{\phi_{a}}
\newcommand{\dmu}{\partial_{\mu}} \newcommand{\umu}{\partial^{\mu}}
\newcommand{\dnu}{\partial_{\nu}} \newcommand{\unu}{\partial^{\nu}}
\newcommand{\di}{\partial_{i}} 
\renewcommand{\dj}{\partial_{j}}  \hoffset =
-2cm \textwidth = 170mm

\begin{document}
\maketitle \abstract{We analyse the structure of minimal-energy solutions of
the  baby Skyrme model for any topological charge $n$; the baby
multi-skyrmions.  Unlike in the (3+1)D nuclear Skyrme model, a potential term
must be present in the (2+1)D baby Skyrme model to ensure stability. The form
of this potential term has a crucial effect on the existence and structure of
baby multi-skyrmions. The simplest holomorphic baby Skyrme model has no known
stable minimal-energy solution for $n$ greater than one. The other baby
Skyrme model studied in the literature possesses non-radially symmetric
minimal-energy configurations that look like `skyrmion lattices' formed by
skyrmions with $n=2$. We discuss a baby Skyrme model with a potential  that
has two vacua. Surprisingly, the minimal-energy solution for every $n$ is
radially-symmetric and the energy grows linearly for large $n$. Further,
these multi-skyrmions are tighter bound, have less energy and the same  large
$r$ behaviour than in the model with one vacuum. We rely on numerical
studies and approximations to test and verify this observation.}

\section{Introduction}

The baby Skyrme model is a modified version of the (2+1)D $S^2$ sigma model.
The addition of a potential and a Skyrme term to the lagrangian ensures
stable solitonic solutions. The Skyrme term has its origin from the nuclear
Skyrme model proposed in \cite{sky:61} and the baby Skyrme model can
therefore be viewed as its (2+1)D analogue.  However, in (2+1) dimensions, a
potential term is necessary in the baby Skyrme models to ensure stability of
skyrmions; this term is optional in the (3+1)D nuclear Skyrme model.

The form of the potential term is largely arbitrary and gives rise to a
multitude of possible baby Skyrme models. In the literature, two specific
models have been studied in great detail (see \cite{pz:93},\cite{psz:95a}).
The simplest holomorphic model does not seem to admit  stable  n-skyrmions
where $n$ is greater than one. (We define an $n$-skyrmion to be the
minimal-energy solution with topological charge $n$.)  And the `baby Skyrme
model'\footnote{it is called baby Skyrme model!\dots we call it old baby
Skyrme model to avoid confusion.}  with a very simple potential possesses
rather beautiful, non-radially symmetric multi-skyrmions. A
new\footnote{first used in \cite{kpz:98}.}  slightly modified potential gives
rise to a remarkably different structure for multi-skyrmions: we call it the
new baby Skyrme model. In fact, we show that the energy density of all
$n$-skyrmions turns out to be radially symmetric configurations, namely rings
of larger and larger radii.  Clearly, the choice of the potential term has a
major impact on the formation of multi-skyrmions and their shape. 

We start with a short introduction to the baby Skyrme models. The earlier
results for the two baby Skyrme models are re-calculated and reviewed in the
light of their multi-skyrmion structure. We present a numerical and
theoretical study of the new baby Skyrme model. We create $n$-skyrmions by
putting $n$ 1-skyrmions in an attractive channel. They form a bound state
which we relax to the minimal-energy state. We find radially-symmetric new
baby multi-skyrmions solutions.  Therefore, we are led to look for static
hedgehog solutions.  Finally, we make some general comments about the
existence and structure of multi-skyrmions depending on the choice of the
potential term.

\section{Baby Skyrme models}

\subsection{Non-trivial Topology \& `Stable' Lagrangian}

Baby Skyrme models\footnote{see review \cite{pz:93}} admit stable field
configurations of finite energy and solitonic nature.  These baby skyrmions
are topological solitons.  Their existence is a consequence of the
non-trivial topology of the mapping of physical space into field space at a
given time t: 
\begin{equation} 
{\cal{M}}: {\cal{S}}^{2} \longrightarrow {\cal S}^{2}.  
\end{equation} 
Here, physical space ${\cal{R}}^2$ is compactified to ${\cal{S}}^{2}$ by
requiring spatial infinity to be equivalent in each direction. This one-point
compactification is necessary to ensure a non-trivial mapping.  The target
manifold (or internal space) is described by a three-dimensional vector $\ph$
with $\ph\cdot\ph=1$.  The non-trivial topology allows to classify maps into
equivalence classes. Each of which has a unique conserved quantity: the
topological charge  
\begin{equation} 
T=\frac{1}{4\pi}\epsilon^{abc}\int dxdy\phi_{a}
\left(\partial_{x}\phi_{b}\right) \left(\partial_{y}\phi_{c}\right)
\end{equation} 
given in integer units. 

Further, stability is ensured by an appropriate choice of lagrangian terms of
field derivatives and a potential. The lagrangian has the form 
\begin{equation}\label{lagrangian} 
L=\partial_{\mu}\ph\cdot\partial^{\mu}\ph
-\theta_{S}\left[(\partial_{\mu}\vec{\phi} \cdot
\partial^{\mu}\vec{\phi})^{2} -(\partial_{\mu}\vec{\phi} \cdot
\partial_{\nu}\vec{\phi}) (\partial^{\mu}\vec{\phi} \cdot
\partial^{\nu}\vec{\phi})\right] - \theta_{V} V(\ph)
\end{equation}
and consists of three terms; from left to right: the sigma model, the Skyrme
and the potential term.  At a classical level, the coefficient of the sigma
model term can always be set to one by re-defining $\theta_S$ and
$\theta_V$. Thus, there are two free parameters in the model.  Each term has
a different scaling behaviour and, together, they ensure stability according
to Derrick's theorem\cite[pages 47-48]{raj}. We require  that the potential
vanishes at infinity for a given vacuum field value; for example
$\vec{\phi}=(0,0,1)$. Care should be taken that the potential term is
invariant under the $SO(2)$ group transformation of $\ph$; this becomes vital
for the use of the hedgehog ansatz. There is a further possibility that the
potential is also zero for other values of the field. Actually, the fact that
the potential vanishes at infinity makes the energy finite and justifies the
one-point compactification of physical space discussed above. 

\subsection{Hedgehog Static Solutions}

The static energy functional density of the baby Skyrme model is
\begin{equation}  
 {\cal E}=\left(\di\ph\cdot\di\ph\right)+ \theta_{S}\left[
\left(\di\ph\cdot\di\ph\right)^{2}-
\left(\di\ph\cdot\dj\ph\right)\left(\di\ph\cdot\dj\ph\right) \right]
+\theta_{V}V(\ph).  
\end{equation}
We look for solutions of the corresponding Euler-Lagrange equation. This is a
very difficult task. The hedgehog ansatz provides a starting  point of our
search for static solutions; in polar coordinates 
\begin{equation} \label{hedgehog} 
\vec{\phi}=\left(\matrix{\sin[f(r)]\sin(n\theta-\chi)\cr\sin[f(r)]
\cos(n\theta-\chi)\cr\cos[f(r)]\cr}\right).   
\end{equation} 
Note that $n$ is a non-zero integer (it is the topological charge as we will
discover later), $\theta$ the polar angle, $\chi$ a phase shift and $f(r)$
the profile function satisfying certain boundary conditions.  The hedgehog
field (\ref{hedgehog}) is chosen, because it is  invariant under the maximal
group of symmetry that leaves the energy functional invariant for non-zero
topological charge (see \cite[page 167]{psz:95a}).  According to the
`Principle of Symmetric Criticality' or `Coleman-Palais theorem'(\cite[pages
72-76]{mrs}), we can search for static solutions invariant under any symmetry
by solving the variational problem for the invariant field. 

The integrated energy density takes the form 
\begin{equation}
E=(4\pi)\frac{1}{2}\int_{0}^{\infty} rdr\left( {f^{'}}^{2} +
n^{2}\frac{\sin^{2}f}{r^{2}}(1+2\theta_{S}{f^{'}}^{2}) +
\theta_{V}\mbox{\~{V}}(f)\right)   
\end{equation}
where $f'=\frac{df}{dr}$.  The energy density depends only on the profile
function $f(r)$: the invariant field. It is independent of the polar angle
and has a radial symmetry. Then, the corresponding Euler-Lagrange equation
with respect to the invariant field $f(r)$ leads to a second-order ODE,
\begin{eqnarray}
  \left(r+\frac{2\theta_{S}n^{2}\sin^{2}f}{r}\right)f^{''} +
  \left(1-\frac{2\theta_{S}n^{2}\sin^{2}f}{r^{2}}+
  \frac{2\theta_{S}n^{2}\sin{f}\cos{f} f{'}}{r} \right)f^{'} \nonumber \\
  -\frac{n^{2}\sin f\cos f}{r}
  -r\frac{\theta_{V}}{2}\frac{d\mbox{\~{V}}(f)}{df} =0, 
\end{eqnarray} 
which we re-write in terms of the second derivative of the profile function:
\begin{equation}
\label{profile} f^{''}={\cal F}(f,f^{'},r). 
\end{equation}
The profile function $f(r)$ is a static solution of the baby Skyrme
model. These static solutions are certainly critical points, but not
necessarily global minima. However, it has been proven that the hedgehog
solution of the nuclear Skyrme model is the minimal-energy solution for
topological charge one (see \cite[pages 80-88]{mrs}). Further, an explicit
hedgehog solution with the topological charge one exists for the holomorphic
baby Skyrme model and has the lowest energy. Therefore, it is  reasonable,
but not proven here, that the hedgehog solution for topological charge one is
the minimal-energy solution.      

The topological charge takes the form 
\begin{equation}
T=-\frac{n}{2}\int_0^{\infty}r dr\left(\frac{f^{'}\sin f}{r}\right)=
 \frac{n}{2}[\cos f(\infty)-\cos f(0)].  
\end{equation}
The boundary conditions for the profile function need to be fixed.  Our value
of the vacuum at infinity is $\vec{\phi}=(0,0,1)$ and we may choose  
\begin{equation}
\label{BCinf} \lim_{r \rightarrow\infty} f(r)=0.  
\end{equation}
Then the value of the profile function at the origin needs to be 
\begin{equation}
\label{BC0} f(0)=m\pi 
\end{equation}
where $m$ is an odd integer. The topological charge is $n$ in integer units.
From now on, we write all other quantities in $4\pi$ units. All $m\neq1$
solutions probably decay into $m=1$ solutions. Thus, in this paper, we
concentrate our attention on solutions corresponding to $m=1$.

\subsection{The Equation of Motion}

A Lagrange multiplier term $\lambda (\vec{\phi}\cdot\vec{\phi}-1)$   needs to
be included in the lagrangian (\ref{lagrangian}) to take care of the $S^2$
constraint (see  \cite[pages 48-58]{raj}). The equations of motion  for each
field component $\phi_{a}$ and $\lambda$ are obtained via the Euler-Lagrange
equation.  Solving for $\lambda$, the equation of motion takes the form
\begin{center}
\begin{eqnarray}
  \dmu\umu\pha-(\ph\cdot\dmu\umu\phi)\pha -2\theta_{S}[ (\dnu\ph\cdot\unu\ph)
  \dmu\umu\pha +(\dmu\unu\ph\cdot\umu\ph)\dnu\pha \nonumber \\
  -(\unu\ph\cdot\umu\ph)\dnu\dmu\pha -(\dnu\unu\ph\cdot\umu\ph)\dmu\pha
  +(\dmu\ph\cdot\umu\ph)(\dnu\ph\cdot\unu\ph)\pha \nonumber \\
  -(\dnu\ph\cdot\dmu\ph)(\unu\ph\cdot\umu\ph)\pha]
  +2\theta_{V}\frac{dV}{d\phi_{3}}(\delta_{a3}-\phi_{a}\phi_{3})=0
\end{eqnarray}
\end{center}
which we re-write in terms of the acceleration of the field $\phi_a$:
\begin{equation}
\label{eom} \partial_{tt}{\phi}_{a}=K_{ab}^{-1} {\cal
  F}_b\left(\ph,\partial_{t}\ph,\partial_{i}\ph\right) 
\end{equation}
with
\begin{equation}
K_{ab}=(1+2\theta_{S}\partial_{i}\ph\cdot\partial_{i}\ph)\delta_{ab}
-2\theta_{S}\partial_{i}\phi_a\partial_{i}\phi_b. 
\end{equation}
We find that the inverse matrix of K exists in an explicit, but rather messy
form. The equation of motion is a second order PDE.

\section{Theoretical Prediction }

Is it possible to predict the general features of the multi-skyrmions? The
main obstacle is the non-linearity of the DEs even in a simpler form like in
the hedgehog ansatz. Nevertheless, at special points i.e.\ the boundaries,
approximations can be made which simplify the DEs. Derrick's theorem allows
us to give further quantitative predictions. Both serve as consistency checks
for our numerical work and help our understanding of the models. 

The value of the field is known at two space locations. Those special points
are $r=0$ and $r=\infty$. In the hedgehog ansatz, the ODE can be approximated
around these points.

\begin{itemize}

\item At the origin, the profile function is approximated as
\begin{equation} \label{f_origin} 
f \simeq \pi+C_{n}r^{n}
\end{equation}
and so
\begin{equation}
f^{'} \simeq nC_{n}r^{n-1} 
\end{equation}
 as long as $\frac{dV(f)}{df}$ tends to zero at this point. We need this
  approximation for the shooting method \cite{profile:94}. Further, the
  energy density at the origin is
\begin{eqnarray}\label{density_origin}
  {\cal E}(0)= & C_1^2(1+C_1^2)+\frac{1}{2}\theta_{V}\mbox{\~V}[\pi]
  &(n=1)\cr {\cal E}(0)= & \frac{1}{2}\theta_{V}\mbox{\~V}[\pi]  &(n\geq2).
\end{eqnarray}
As one would expect, the energy density of the new baby Skyrme model is zero,
because there is a further vacuum at the origin;$\mbox{\~V}[\pi]=0$. It is
non-zero only in the topological sector one. Clearly, if the hedgehog
solutions are the minimal-energy solutions, the new baby $n$-skyrmions are
ring configurations.  However, in the  case of the old or simplest
holomorphic  baby Skyrme model, the energy density is always non-zero for any
static hedgehog solution. These hedgehog solutions do not seem to minimise
the energy as well as in the new baby model. Our numerical results confirm
this.

\item At large r, the ODE reduces to
\begin{equation}
\label{larger} f^{''}+\frac{1}{r}f^{'}-\frac{n^2}{r^2}f
  -\frac{\theta_{V}}{2}\frac{d\mbox{\~V}(f)}{df} {\big \vert}_{\mbox{large
 r}}=0. 
\end{equation}
  The last term, arising from the potential, can be neglected for some
  potentials in a consistent way, because it is small compared to the other
  terms. This is the case for the holomorphic model where this term   is of
  order $f^7$. However, the term has to be included for the old and new baby
  Skyrme model and gives $-\frac{\theta_{V}}{2}f$ or $-\theta_{V}f$
  respectively.  Looking at large r, the old and the new baby skyrmions
  behave in the same way\footnote{we just re-define $\theta_{V}$ for one of
  the models.}. Actually, the DE is that of a static Klein-Gordon field with
  a radially symmetric form where $\theta_{V}$ plays the role of the meson
  mass. As discussed above, the real difference lies in the small $r$ and
  medium $r$ region.  For the new baby Skyrme model, the equation
  (\ref{larger}) gives
\begin{equation}
 f^{''}+\frac{1}{r}f^{'}-f(\frac{n^2}{r^2}+\theta_{V})=0.  
\end{equation}
The coefficient of the potential term is present here. The potential
localises the skyrmion exponentially. Solving for appropriate boundary
conditions, the profile function decays exponentially
\begin{equation}
f(r)\longrightarrow\frac{1}{\sqrt{\theta_{V}r}}\exp(-\theta_{V} r).  
\end{equation}

\end{itemize}

The Derrick theorem\footnote{also called Hobart-Derrick theorem} (see
 \cite[pages 52-54]{mrs}) provides a necessary but not sufficient condition
 for the existence of stable solutions. Under a simple scale transformation
 $r\rightarrow\lambda r$, the total energy changes to a function of $\lambda$
 and the non-scaled energies of the three terms:
\begin{equation} 
\label{E_lambda} 
E[\tilde f(\lambda r)]=E_{\sigma}+\lambda^2\theta_S E_s +\lambda^{-2}
\theta_V E_V.   
\end{equation}
The sigma term is scale invariant.  The derivative of the energy with respect
 to $\lambda$ at $\lambda=1$ has to be zero if a stable solution exists. This
 implies 
\begin{equation}
\theta_S E_s = \theta_V E_V.   
\end{equation}
Our numerical results have to fulfill this condition.  Further
(\ref{E_lambda}) suggests that the scaling effect can be  un-done by
redefining $\theta_S$ to $\lambda^2\theta_S$ and $\theta_V$ to
$\lambda^{-2}\theta_V$. In fact, writing the DE (\ref{profile}) in terms of
$\tilde f(\tilde r,\tilde\theta_S,\tilde\theta_V)$ and using $\tilde
r=\lambda r$, we find 
\begin{equation}
\tilde f(\lambda r,\lambda^{-2}\theta_S,\lambda^2
\theta_V)=f(r,\theta_S,\theta_V).
\end{equation}
Substituting $\tilde f$ into the energy functional gives exactly the same
energy as $f$ does. If two models with coefficients $\theta_S$ and $\theta_V$
respectively. $\tilde\theta_S$ and $\tilde\theta_V$ satisfy
\begin{equation}
\label{equal}\theta_S\theta_V=\tilde\theta_S\tilde\theta_V,
\end{equation}
then their stable solutions have the same energy. This is a further check on
our numerical results.
    
\section{Numerical Techniques}

The baby Skyrme model is a non-integrable system and explicit solutions to
its resulting differential equations are nearly impossible to find. Numerical
methods are the only way forward. We need (\ref{eom}) for the time-evolution
and relaxation of an initial configuration and use (\ref{profile}) to find
static hedgehog solutions. These DEs are re-written as sets of two first
order DEs. We discretise DEs by restricting our function to values at lattice
points and by reducing the derivatives to finite differences.\footnote{we use
the 9-point laplacian}(as explained in \cite{pz:98}, see also \cite{num}). We
take the time step to be half  the lattice spacing: $\delta
t=\frac{1}{2}\delta x$. We use fixed boundary conditions i.e.\ we set the
derivatives to zero at the boundary. We check our numerical results via
quantities conserved in the continuum limit and by changing lattice spacing
and number of points. Moreover, we compare them with theoretical predictions.

\paragraph{Looking for static hedgehog solutions}
The static hedgehog solutions of (\ref{profile}) are found by the shooting
method using the 4th order Runge-Kutta integration and the boundary
conditions (\ref{BCinf}) and (\ref{BC0}). Alternatively, one can use a
relaxation technique like the Gauss-Seidel over-relaxation (\cite{num})
applied to an initial configuration with the same boundary conditions.  

\paragraph{Looking for multi-skyrmions}

We construct $n$-skyrmions by relaxing an initial set-up of n 1-skyrmions
with relative phase shift of $\frac{\pi}{n}$. Using the dipole picture
developed by Piette et al. \cite{psz:95a}, two old (or new) baby 1-skyrmions
attract each other for a non-zero value of the relative phase; phase shift of
$\frac{\pi}{2}$ for maximal attraction. A circular set-up is crucial as they
maximally attract each other and 1-skyrmions do not form several states that
repel each other. One possible objection to a circular set-up is its apparent
discrete symmetry, the cyclic group $Z_n$. However, the discretised PDE on a
finite square lattice is not invariant under $Z_n$ as we impose boundary
conditions. Further, the linear superposition is only an approximate solution
of the model and the 1-skyrmions used are produced from the hedgehog ansatz.
In this sense, lattice effects and small integration errors even provide
useful small perturbations. We have run simulations for non-circular set-ups,
but either the 1-skyrmions take longer to fuse together or they fuse into
many bound-states and repel each other. We run our simulations on grids with
$200^2$ or $300^2$ lattice points and the lattice was $\delta x=0.1 \mbox{ or
} 0.05$. However, for large topological charge, we need larger grids and the
relaxation takes a long time. The corresponding hedgehog solution as an
initial set-up usually works well and is faster, but biased due to its large
symmetry group.  The time-evolution of an initial configuration is determined
by the equation of motion (\ref{eom}). We are using the 4th order Runge-Kutta
method to evolve the initial set-up and correction techniques to keep the
errors small. We relax i.e.\ take out kinetic energy by using a damping (or
friction) term.

\paragraph{Initial set-up:} 
The initial field configuration is a linear superposition of static solutions
 with or without initial velocity; typically we use a circular set-up of n
 1-skyrmions with a $\frac{\pi}{n}$ phase-shift between each other (for
 maximal attraction).  The superposition is justified, because the profile
 function  decays exponentially. The superposition is done in the complex
 field formalism i.e.\ where $W$ is the stereographic projection of the $\ph$
 field of $S^2$ (see \cite{pz:93}). We use the profile function of a static
 solution (typically of topological charge one) to obtain  
\begin{equation}
  W=\tan\left(\frac{f(r)}{2}\right)e^{-in\theta}.
\end{equation}
This equation holds in the rest frame of a static skyrmion solution centred
around its origin and $\frac{dW}{dt}=0$. We may introduce moving solutions by
switching to a different frame of reference. This can be done by performing a
Lorentz boost on the rest frame of a given $W$, because $W$ is a Lorentz
scalar. Now it is  easy to construct a linear superposition of individual,
moving or not, baby skyrmion solutions $W_{\alpha}$ by 
\begin{equation}
W(x,y)=\sum_{\alpha} W_{\alpha}(x-x_{\alpha},y-y_{\alpha})
\end{equation}
where ($x_{\alpha}$,$y_{\alpha}$) is the location of the centre of the
{$\alpha$}th skyrmion.   It is  important that the different skyrmions are
not too close to each other.  Finally, the complex field is re-written in
terms of the field $\ph$ and its derivative. 

\paragraph{Correction techniques:} 
The integration method introduces small errors which eventually add up.   In
terms of the $S^2$ constraint, this corresponds to the field leaving the
two-sphere. Hence, we need to project the field back onto the sphere.  The
simplest and sufficient projections are
\begin{equation}
\phi_{a}\longrightarrow \frac{\phi_{a}}{\sqrt{\ph\cdot\ph}} 
\end{equation}
and
\begin{equation}
\partial_{t}\phi_{a}\longrightarrow\partial_{t}\phi_{a}-
\frac{\partial_{t}\ph\cdot\ph}{\ph\cdot\ph}\phi_{a}.
\end{equation}
Of course, the space derivatives may also be corrected. See \cite{pz:98} for
further discussions.

\paragraph{Relaxation technique:} 
A damping term in the equation of motion will gradually take the kinetic
energy out of the system. The equation (\ref{eom}) changes to
\begin{equation}
\partial_{tt}{\phi}_{a}=K_{ab}^{-1} {\cal
  F}_b\left(\ph,\partial_{t}\ph,\partial_{i}\ph\right)
  -\gamma\partial_{t}\phi_a
\end{equation}
where $\gamma$ is the damping coefficient. We set $\gamma$ to 0.1, but most
values will do as long as they are not too large. Another approach would be
to absorb the outwards travelling kinetic energy waves in the boundary region.

\section{The Different Models}

So far, the literature on baby skyrmions reports work on the holomorphic
model with $V=(1+\phi_{3})^4$ and the old baby Skyrme model with
$V=1-\phi_3$. There are no stable multi-skyrmions found in the holomorphic
model.  However, the old baby Skyrme model possesses non-radially symmetric
minimal-energy solutions. We will show that the new baby Skyrme model with
$V=1-\phi_3^2$ has radially symmetric multi-skyrmions.

\subsection{Holomorphic Model}

The simplest holomorphic model has the potential $V=(1+\phi_{3})^4$ and is
the first baby Skyrme model studied in the literature (\cite{lpz:90},
\cite{sut:91}, \cite{pz:93}). We have re-done the calculations and agree with
the literature. This agreement provides a check on our numerical methods.

The holomorphic potential is unique in the sense that its model admits an
explicit solution for a skyrmion with topological charge one (we call it a
1-skyrmion). To leading order, the asymptotic behaviour  does not depend on
the potential. The skyrmion is polynomially localised. The force between two
holomorphic skyrmions is always repulsive. This repulsion can be overcome by
sending the two 1-skyrmions against each other at a sufficiently high
speed. Above a critical value, they overlap and form a intermediate
state. However, this state is not stable and the two 1-skyrmions scatter at
90 degrees. No multi-skyrmions are known to exist in this model.

\subsection{Old Baby Skyrme Model}

The old baby Skyrme model has been extensively studied in \cite{psz:95a} and
\cite{psz:95b}. The potential $V=1-\phi_3$ gives rise to very structured
multi-skyrmions. The configurations are crystal-like in the sense that its
building block is a 2-skyrmion. We have re-done Piette et al.'s computations
for multi-skyrmions and confirm their results. We use their coefficients
i.e.\ add a factor $\frac{1}{2}$ to the sigma model term, $\theta_{S}=0.25$
and $\theta_{V}=0.1$ [see (\ref{lagrangian})].

\paragraph{Multi-skyrmions}
 We construct a 2-skyrmion by sending two 1-skyrmions against each other.
Put close to each other, they merge into  an oscillating bound-state (see
next section for picture) which leads to the stable radially symmetric
2-skyrmion by numerical relaxation. We have checked its stability by evolving
it in time without relaxation. We repeat the procedure and find that all
higher $n$-skyrmions are not radially symmetric.  We  extend Piette et al.'s
work to $n=7$, $n=8$ and $n=9$ to make sure that the $n$-skyrmions are
`skyrmion crystals' formed by 2-skyrmions. In this paper, we only present the
formation of an 8-skyrmion (see figure~\ref{baby_merge8}). We put eight
1-skyrmions on a circle with a phase shift of $\frac{\pi}{n}$ between
neighbouring 1-skyrmions. The initial configuration of eight 1-skyrmions is
time-evolved and relaxed. The system starts moving to four 2-skyrmions which
re-arrange themselves.  Slowly, the system moves towards a stable
configuration, the 8-skyrmion. The building block of this crystal-like
structure is the 2-skyrmion. 
\begin{figure}
\begin{center}
\caption{Contour plots of energy density: Eight 1-skyrmions merge together (pictures 1-4)}
\label{baby_merge8}
\includegraphics[angle=0,scale=1,width=0.45\textwidth]{baby_merge8_a.epsi}
\caption{Contour plots of energy density: Bound state relaxes into an 
 8-skyrmion (pictures 5-8)}
\includegraphics[angle=0,scale=1,width=0.45\textwidth]{baby_merge8_b.epsi}
\end{center}
\end{figure}

\paragraph{Hedgehog Ansatz} 
Solutions of the hedgehog ansatz can be found numerically by solving equation
(\ref{profile}) via the shooting method. As seen above, only the $n=1$ and
$n=2$ hedgehog configurations are global minima. The 1-skyrmion has a hill
shape and is exponentially localised.  Unlike in the holomorphic model, the
asymptotic behaviour of the profile function does depend on the potential
term to leading order. The 2-skyrmion is a ring-like configuration.  A $n=3$
static hedgehog solution exists, but does not have the lowest energy in its
topological sector. A time-evolution shows that it is  unstable and relaxes
to the non-radially symmetric 3-skyrmion.
\begin{table}\label{table:old}
\begin{center}
\begin{tabular}{|c||c|c|c|c|}\hline
  Charge & Energy & Energy per skyrmion & Break-up modes & Ionisation Energy
  \cr\hline 1 & 1.549 & 1.549 & - & - \cr\hline 2 & 2.907 & 1.454 & $1+1$ &
  0.191 \cr\hline 3 & 4.379 & 1.460 & $2+1$ & 0.077 \cr\hline 4 & 5.800 &
  1.450 & $2+2$ & 0.014 \cr & & & $3+1$ & 0.128 \cr\hline 5 & 7.282 & 1.456 &
  $3+2$ & 0.005 \cr & & & $4+1$ & 0.068 \cr\hline 6 & 8.693 & 1.449 & $4+2$ &
  0.015 \cr & & & $3+3$ & 0.066 \cr & & & $5+1$ & 0.138 \cr\hline
\end{tabular}
\end{center}
\caption{Multi-skyrmions of the old baby Skyrme model}
\end{table}

\paragraph{Energies}
Table 1 reproduces Piette et al.'s results and we will compare them to the
 new baby model's. The ionisation energy $E_{kl}$ is defined as the energy
 you have to add to a n-skyrmion to break it up into a k-skyrmion and a
 l-skyrmion:  
\begin{equation}
E_{kl}=E_n-(E_k+E_l). 
\end{equation} 
As $n$ increases, the ionisation energy decreases (but not monotonically) and
the $n$-skyrmions become less bound. The 2-skyrmion emission is the
energetically most favoura\-ble break-up mode. An emission of a 2-skyrmion
takes the smallest amount of kinetic energy to break up an 8-skyrmion.  Thus,
it is  justified to think about a $n$-skyrmion as a collection of 2-skyrmions
bound together. 

 Note that the data comes from the full time-evolution.  Using the hedgehog
ansatz leads to slightly different, more accurate, values.  There the energy
of a 1-skyrmion is 1.564 and the energy of a 2-skyrmion is 2.936. This effect
is due to the finite lattice and the fact that we can use more lattice points
in the hedgehog ansatz.

\subsection{New Baby Skyrme Model}

The new baby Skyrme model\footnote{see also \cite{kpz:98}} exhibits a
completely different structure for multi-skyrmions. In fact, the
multi-skyrmions are ring-like configurations; their radii being proportional
to their topological charge. The form of the potential is
\begin{equation}
V=1-\phi_3^2=(1-\phi_3)(1+\phi_3). 
\end{equation}
The potential has two vacua; for $\phi_3=1$ and $\phi_3=-1$. At infinity, the
old and the new baby Skyrme models have the same vacuum $\phi_3=1$ and behave
in the same way. They only differ for small $r$.  Another important fact is
that the lagrangian of the new baby Skyrme model is invariant under $ \ph
\longrightarrow -\ph$.

\subsubsection*{Numerical Results}

\paragraph{Looking for multi-skyrmions}

\begin{figure}
\begin{center}
\caption{Contour plot of energy density: Formation of 2-skyrmion}
\label{new_baby_scat_2}
\includegraphics[angle=0,scale=1,width=0.45\textwidth]{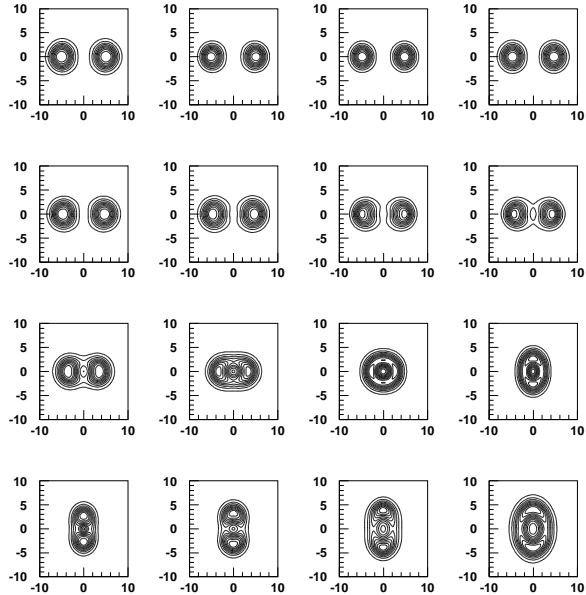}
\end{center}
\end{figure}

\begin{figure}
\begin{center}
\caption{Contour Plot of Energy Density: Formation of 5-skyrmion}
\label{new_baby_merge}
\includegraphics[angle=0,scale=1,width=0.4\textwidth]{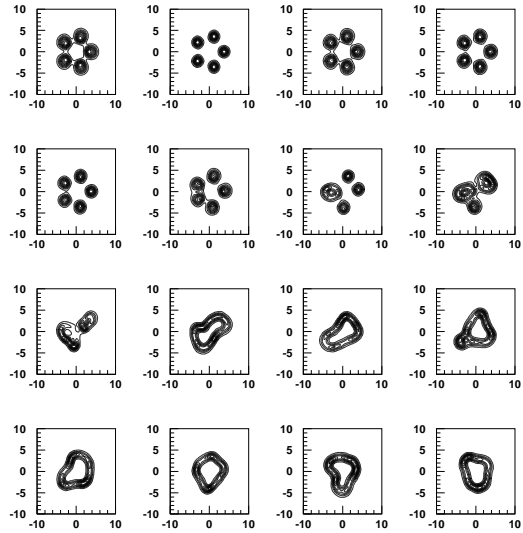}
\end{center}
\end{figure}

\begin{figure}
\begin{center}
\caption{Contour Plot of Energy Density: Rings of multi-skyrmions from $n=2$ to $n=5$}
\includegraphics[angle=0,scale=1,width=0.4\textwidth]{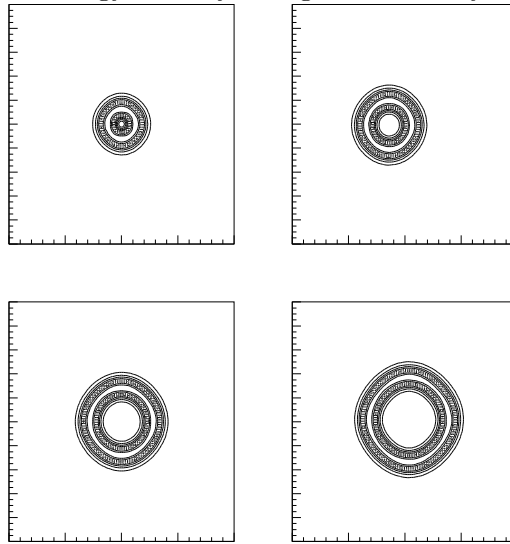}
\label{rings}
\end{center}
\end{figure}

First, two new baby 1-skyrmions scatter in the same way as the old baby
skyrmions do. Figure~\ref{new_baby_scat_2} shows how the two 1-skyrmions
attract each other, form a bound-state, scatter away at 90 degrees, get
slowed down by their mutual attraction, attract each other again and so on.
This oscillating but stable bound-state is an excited state of the 2-skyrmion
solution. Taking out the kinetic energy, the bound-state relaxes to the
2-skyrmion; a ring.
We have looked at all higher n-skyrmions.  Figure~\ref{new_baby_merge}
presents the results of one of our simulations: the formation of a
5-skyrmion.  The skyrmions attract each other and merge into intermediate
states.  Relaxation takes out the kinetic energy and the unstable
intermediate states merge together. They form an irregular ring configuration
that moves like a vibrating closed string.  Slowly, the configuration settles
down to a radially symmetric form due to the loss of kinetic energy.
Figure~\ref{rings} shows the final configuration of multi-skyrmions from
charge two to five. Actually, the larger rings are slightly deformed. This
effect is due to boundary effects and reduced by using larger grids.  Thus,
we have convinced ourself of the radial symmetry of $n$-skyrmions.

\paragraph*{In the hedgehog ansatz}

Clearly, the $n$-skyrmions are radially symmetric and we can study them in
the hedgehog ansatz. Numerically speaking, the problem is reduced to one
dimension and, effectively, we can take as many lattice points as we want. We
want to compare the multi-skyrmions to those of the old baby Skyrme model. We
add a factor of $\frac{1}{2}$ to the sigma term. The coefficient $\theta_{V}$
is set  to half of the value of $\theta_{V}$ in the old baby Skyrme model:
$\theta_{V}=0.05$. Now, the old baby Skyrme model has exactly the same large
r behaviour i.e same pion mass. Then, we set $\theta_{S}$ so that the energy
of the new baby 1-skyrmion is now approximately the same as the old baby
1-skyrmion in the hedgehog ansatz: $\theta_{S}=0.44365$. This convention puts
both models on  an equal footing. Note that there is a certain ambiguity
about the choice of the coefficients. The energy is a function of
$\theta_{S}\theta_{V}$  and a compensating change of both coefficients gives
the same energy.  

\begin{figure}
\caption{The energy density of new baby skyrmions up to charge 10}
\label{new_baby_energies}
\begin{center}
\includegraphics[angle=-90,scale=1,width=0.5\textwidth]{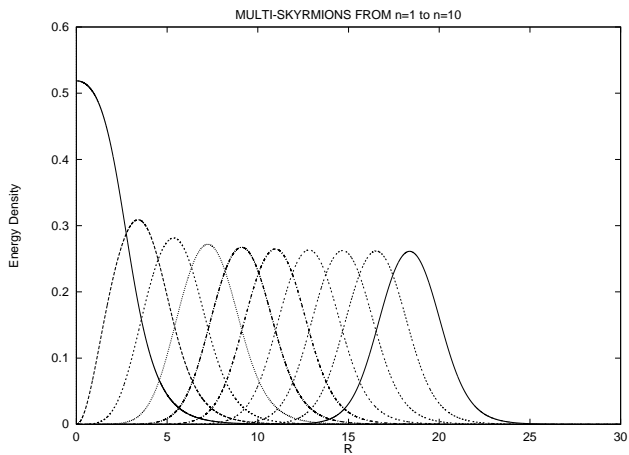}
\end{center}
\end{figure}

Again, we find the hedgehog solutions by solving (\ref{profile}) using the
shooting method. Figure~\ref{new_baby_energies} shows all solutions up to
topological charge 10.  The higher the charge the more difficult it becomes
to find solutions numerically. The shooting method becomes more and more
sensitive to the numerically determined value of its derivative at the origin
i.e. $C_{10}\approx10^{-10}$(see~\ref{f_origin}).
The numerical results are very interesting. The peaks of the energy density
of rings converge to an asymptotic height and their position shifts by an
asymptotically constant amount.  This observation deserves some further
understanding.

\paragraph{Relation between energy peak and its location}
Our numerical results show that the profile function $f$ at the position of
the energy density peak, $r=d(n)$, approaches the value $\frac{\pi}{2}$ for
large $n$. The energy density at this point reduces to
\begin{equation}{\cal E}
[d(n),n]={f_d^{'}}^2 + \frac{n^2}{d^2}(1+2\theta_{S}{f_d^{'}}^2)+\theta_{V}.  
\end{equation}
Its value depends on the derivative of the profile function at $d(n)$ and the
ratio between $n$ and $d(n)$. Now, figure~\ref{new_baby_energies} shows that
the height of the energy density of the peak is approximately a constant for
large $n$ i.e.
\begin{equation}
\lim_{n\rightarrow\infty} {\cal E}[d(n),n]=constant. 
\end{equation}
The larger the topological charge the more the multi-skyrmion approaches the
peak and shape of the `asymptotic multi-skyrmion'. Using this empirical
knowledge leads us to conclude that, in the large $n$ limit,
\begin{equation}
\lim_{n\rightarrow\infty} {f_d^{'}}^2=\alpha^2
\end{equation}
and
\begin{equation}
\lim_{n\rightarrow\infty} \frac{n^2}{d(n)^2}=\beta^2,
\end{equation}
where $\alpha$ and $\beta$ are constants. And, the peak of the ring shifts by
a fixed amount from a skyrmion of charge $n$ to one of charge $n+1$ i.e.
\begin{equation}
\label{npropd} d(n)=\frac{1}{\beta} n.  
\end{equation}
This relation agrees with our numerical work and provides a  good consistency
check. The energy density becomes
\begin{equation}
{\cal E}[d(\infty)]=\alpha^2 +\beta^2(1+2\theta_S\alpha^2)+\theta_{V}. 
\end{equation}
The shape of the $n$-skyrmions approaches
that of the `asymptotic multi-skyrmion'.  We can approximate the
configurations by a finite box of height ${\cal E}_d$ around the point
$r=d(n)$. This gives us the dependence of the total energy on the topological
charge i.e. 
\begin{equation}
E=\int_{d(n)-a}^{d(n)+a}dr r {\cal E}_d=2{\cal E}_d a d(n) =(2{\cal E}_d
    a\beta^2)n 
\end{equation}
using (\ref{npropd}). Asymptotically, the total energy grows linearly with
$n$.  Linear dependence suggests that radially symmetric solutions are global
minima. The system does not switch to less symmetric  configurations, because
they do not have linear dependence on $n$.

\paragraph{Energies} 
Table 2 shows our numerical results for the energies of the multi-skyrmions
  up to $n=6$. First, the 2-skyrmion has a lower energy than an old baby
  2-skyrmion, but looks exactly the same at large $r$.  The multi-skyrmions
  do not break up via a 2-skyrmion emission (see last section) but  into two
  similar configurations ie.  5$\rightarrow3+2$ or 6$\rightarrow3+3$. This
  can be seen from the ionisation energy.
\begin{table}\label{table:new}
\begin{center}
\begin{tabular}{|c||c|c|c|c|}\hline
  Charge & Energy & Energy per skyrmion & Break-up modes & Ionisation Energy
  \cr\hline 1 & 1.564 & 1.564 & - & -  \cr\hline 2 & 2.809 & 1.405 & $1+1$ &
  0.319 \cr\hline 3 & 4.112 & 1.371 & $2+1$ & 0.262 \cr & & & $1+1+1$ & 0.580
  \cr\hline 4 & 5.433 & 1.358 & $2+2$ & 0.186 \cr & & & $3+1$ & 0.243
  \cr\hline 5 & 6.761 & 1.352 & $3+2$ & 0.160 \cr & & & $4+1$ & 0.235
  \cr\hline 6 & 8.094 & 1.349 & $3+3$ & 0.130 \cr & & & $4+2$ & 0.148
  \cr\hline
\end{tabular}
\end{center}
\caption{Multi-skyrmions of the new baby Skyrme model}
\end{table}

Further the ionisation energy and the energy per skyrmion decreases to an
asymptotic value for large $n$ (unlike the old baby multi-skyrmions). The
monotonic decrease of the ionisation energy shows that the large
$n$-skyrmions become less stable: a  smaller addition of kinetic energy can
break them apart.  Nevertheless, they are much tighter bound and more stable
than their old baby analogues.

\section{Summary and Open Questions}

Clearly, the choice of the potential term has a crucial effect on the
structure of multi-skyrmions. The comparison between the new and the old baby
Skyrme model has proved to be very interesting.  Both models have the same
asymptotic behaviour, but possess completely different multi-skyrmion
structures.  The new baby Skyrme model has radially symmetric minimal-energy
solutions  for all topological charges whereas the old baby multi-skyrmions
are `skyrmion lattices' formed by 2-skyrmions. New baby multi-skyrmions are
tighter bound and have less energy than their old baby analogues. 

We have backed up our numerical results by  monitoring conserved quantities
(like energy, topological charge, $S^2$ constraints),  comparing with
approximations for small and large r, checking the relation between energy
peaks and their position and verifying conditions imposed by  Derrick's
theorem. 

Obviously, a general framework that predicts the structure of multi-skyrmions
for a given potential is desirable. However, we have not been able to achieve
this goal. Rather, we state some empirical laws derived from our numerical
experiments. 

\begin{description}
\item{\bf Existence of multi-skyrmions.} Baby Skyrme models seem to admit
  stable multi-skyrmions only if the force between two 1-skyrmions can be
  attractive. Two 1-skyrmions overlap, form an intermediate state and scatter
  at 90 degrees.  Only an attractive force between them can overcome the
  energy due to the scattering and lead to a bound state.  A related
  observation leads us to the conjecture that if the asymptotic behaviour
  does not depend on the potential coefficient, the force between two
  1-skyrmions is repulsive; see the holomorphic model.

\item{\bf Structure of multi-skyrmions.} Clearly, the potential shapes the
  structure of multi-skyrmions. Unfortunately, we can only show this fact by
  our numerical results. The existence of more than one vacuum seems to be
  crucial to the radial symmetric shape of the new baby $n$-skyrmions. It
  might well be possible to prove that potentials with more than one vacuum
  lead to radially symmetric multi-skyrmions. However, we were not able to
  back  such claims by a theoretical study. 

\end{description}

To    conclude,  some   interesting  questions   arise   from  the  study  of
multi-skyrmions in  the   baby Skyrme models    and  are worth  investigating
further. 

\begin{itemize}
\item The choice for potential terms is largely arbitrary. The study of other
potentials and in particular those with multiple  vacua could help to clarify
the issues surrounding     the     existence     and  structure   of
multi-skyrmions.  Further, on  one hand, potentials  with  multiple vacua can
still have the same large distance behaviour like  old baby skyrmions. On the
other hand, their multi-skyrmion  solutions should have rather exotic shapes,
because the dynamics  tries to have  multiple vacua. Unlike the  conventional
smooth lump 1-skyrmion, their 1-skyrmion lumps may have riddles in them. Or,
one might like to rephrase our question and ask whether multi-vacua models
have circular domain walls, too. 

\item  Of course, the application of   the baby Skyrme   model in the quantum
  regime requires an appropriate  quantization scheme. The radially symmetric
  new baby multi-skyrmion  solutions simplify this task.  Unlike the old baby
  Skyrme model,  the mass  correction  of {\it all multi-skyrmions} \/ can be
  calculated numerically without  considerable computer power.  The next step
  should be the quantization of the baby Skyrme model. 
\item  Are there  any   applications? If  there   are 90  degrees  scattering
  phenomena observed  in experiments, then the   analysis of bound-states can
  lead to the determination of the potential responsible for these phenomena. 
\item  The   choice of the    potential  crucially shapes   the  structure of
  multi-skyrmions.  This is probably  not true  in  the nuclear Skyrme model,
  because   the sigma term is  not   scale invariant.  Often,   an `old  baby
  Skyrme'-type potential is included to have a pion mass and exponential
  decay.  There is no  argument why the  potential term  used cannot  have
  two vacua.  What happens in a nuclear Skyrme model with a  `new baby Skyrme
  potential'?  Is the energy of the $n$-skyrmions lower? 
\end{itemize}

It is possible to view the ring-like multi-skyrmions of the new baby Skyrme
model as circular domain walls separating the two vacua. This is an
interesting and conceptually clearer viewpoint. Indeed, at the location of
the peak of the energy density, the field switches from one vacuum value to
the other. These domain walls also have a topological charge.  

I would like to thank Wojtek Zakrzewski and Bernard Piette for their support.
I also used  some subroutines   written by Bernard   Piette.  Further, I  am
very grateful to  the referees for    very detailed and  constructive
feedback.  I acknowledge   the  receipt   of a    Durham Research   Award and
a  `bourse post-universitaire' from the government of Luxembourg.

\bibliographystyle{plain} \bibliography{paper_ref}
\end{document}